\documentstyle[aps,amsmath,amsfonts,epic,eepic,rotating]{revtex}

\newcommand{\ket}[1]{{| #1 \rangle}}
\newcommand{\bra}[1]{{\langle #1 |}}
\newcommand{\normsqr}[1]{{\langle #1 | #1 \rangle}}
\newcommand{\smallspace}{{\mskip 2mu minus 1mu}}

\newcommand{\Co}{{\mathbb C}}

\newlength{\coslength}
\newlength{\msinlength}
\def\mylangle{\texttt <}
\def\myrangle{\texttt >}
\newcommand{\myurl}[1]{\mylangle http:/$\mskip -2.5mu$/{#1}\myrangle}

\title{On~Arbitrary Phases in Quantum\\ Amplitude Amplification}
\author{Peter H{\o}yer\thanks{\,email:
\mbox{\texttt{hoyer}\textbf{\char"40}\texttt{brics.dk}}.}\\
BRICS\thanks{\,Basic Research in Computer Science,
Centre of the Danish National Research Foundation.}\,,
Department of Computer Science, University of Aarhus,\\
Ny~Munkegade, Bldg.~540, DK-8000 Aarhus~C, Denmark.}

\begin{document}
\draft
\maketitle

\begin{abstract}
We~consider the use of arbitrary phases in quantum amplitude 
amplification which is a genera\-li\-zation of quantum searching.
We~prove that the phase condition in amplitude amplification
is given by $\tan(\varphi/2)=\tan(\phi/2)(1-2a)$, 
where $\phi$ and~$\varphi$ are the phases used 
and where $a$ is the success probability of the given algorithm.
Thus the choice of phases depends nontrivially and nonlinearly
on the success probability.
Utilizing this condition, we give methods for constructing 
quantum algorithms that succeed with certainty
and for implementing arbitrary rotations.
We~also conclude that phase errors of order up to~$\frac{1}{\sqrt{a}}$ 
can be tolerated in amplitude amplification.
\end{abstract}



\section{Introduction}
\label{sec:intro}

Most quantum algorithms developed so far are based on 
two techniques: 
Quantum Fourier Transforms and Amplitude Amplification.  
The latter technique is a generalization of Grover's quantum 
algorithm for searching an unordered database~\cite{Grover97}, 
and it allows a quadratic speedup over any classical algorithm
for many computational problems.
Since amplitude amplification is fundamental for quantum
algorithms it has received a great deal of attention.
This includes a study of its robustness to errors and 
modifications.
In~particular, the effects of using arbitrary phases in 
amplitude amplification have been studied in a sequence of papers 
by Long \emph{et~al.}~\cite{LZLN99,LLZN99,LTLZY99,LLZT00}.

In~this paper, we also consider the question of when arbitrary
phases can be utilized in amplitude amplification.
Our results complement the results of Long \emph{et~al.} whom
are primarily interested in the question
of how large phase errors we can tolerate and still obtain
a quantum algorithm for searching that succeeds with \emph{high
probability}.
We~are primarily interested in the question of what restrictions
we need to put on the two angles used in amplitude amplification
and still obtain quantum algorithms that succeed 
with \emph{certainty}.

{To}~illustrate our results,
consider a database of size~$N$ with a unique marked element.
In~each iteration of Grover's algorithm, we rotate the 
phase of some states by angles $\phi$ and~$\varphi$, respectively.
The main result in~\cite{LZLN99,LLZN99,LTLZY99,LLZT00} is
that if the two angles differ by at most $\frac{c}{\sqrt{N}}$ 
for some appropriate constant~$c$, that is, 
if $|\varphi-\phi| \leq \frac{c}{\sqrt{N}}$, 
then we can still find the marked element with high probability
using only $\Theta(\sqrt{N})$ iterations.
Our main result is that if the two angles satisfy the equation
$\tan(\varphi/2) = \tan(\phi/2) (1-\frac{2}{N})$ then
we can find the marked element with certainty using
only $\Theta(\sqrt{N})$ iterations.
Together with the result of Long \emph{et~al.}, this
provides a description of the use of arbitrary phases
for bounded-error and exact quantum algorithms.
It~is possible to rederive the main result of Long \emph{et~al.} 
from our results by considering the case $\phi=\varphi$ an
approximation to the perfect case 
$\tan(\varphi/2) = \tan(\phi/2) (1-\frac{2}{N})$.
(See Secs.~\mbox{\ref{sec:arbitrary}--\ref{sec:phasecondition}}
below for rigorous statements.)

We~thus prove that there is phase condition in
amplitude amplification,
and that this condition is not that the phases are equal, 
but that they satisfy the trigonometric equation mentioned above.
We~believe that our approach is intuitive
and that it yields short and straightforward proofs.

\section{Amplitude amplification}
\label{sec:ampamp}
Amplitude amplification is a generalization of Grover's 
quantum searching algorithm that allows a speed up of 
many classical algorithms.  
The heart of amplitude amplification is 
an operator~${\mathbf Q}$ defined similarly to 
the operator used in Grover's algorithm~\cite{Grover97}.
We~refer the reader to~\cite{BHMT00} and the references therein
for a throughly introduction to amplitude amplification.
Here we give only a concise description of all the objects 
we require.

Let ${\mathcal H}$ be Hilbert space of dimension~$N$ 
and let $\{\ket{0},\ldots,\ket{N-1}\}$ be an orthonormal basis
for~${\mathcal H}$.
Let ${\mathcal A}$ be a unitary operator on~$\mathcal H$.
We~may think of ${\mathcal A}$ as a quantum algorithm that uses
no measurements.
Let $\chi: \{0,\ldots,N-1\} \rightarrow \{0,1\}$ be a
Boolean function.
We~say that a basis state $\ket{x}$ is \emph{good\/} 
if $\chi(x)=1$, 
and otherwise we say that $\ket{x}$ is \emph{bad}.
Given two angles $0 \leq \phi, \varphi < 2 \pi$, define
\begin{equation}
\label{eq:defq}
{\mathbf Q} = {\mathbf Q}({\mathcal A},\chi,\phi,\varphi)
  = - {\mathcal A} \smallspace {\mathbf S}_0(\phi) \smallspace
    {\mathcal A}^{-1} \smallspace {\mathbf S}_\chi(\varphi).
\end{equation}
Here, the operator ${\mathbf S}_\chi(\varphi)$ 
conditionally changes the phase of the amplitudes 
of the good states,
\begin{equation}
\ket{x} \;\longmapsto\; 
\begin{cases}
e^{\imath \varphi} \ket{x} & \text{if $\chi(x)=1$}\\
\hphantom{e^{\imath \varphi}} \ket{x}
  & \text{if $\chi(x)=0$.}\end{cases}
\end{equation}
Similarly, the operator ${\mathbf S}_0(\phi)$
multiplies the amplitude by a factor of~$e^{\imath \phi}$
if and only if the state is the zero state~$\ket{0}$.
Here and elsewhere we use~$\imath$ to denote the 
principal square root of~$-1$.

Let $\ket{\Psi} = {\mathcal A}\ket{0}$ denote the superposition
obtained by applying algorithm ${\mathcal A}$ on the initial 
state~$\ket{0}$. 
Let $\ket{\Psi_1} = {\mathbf P}_{\text{good}} \ket{\Psi}$
where ${\mathbf P}_{\text{good}} = \sum_{x:\chi(x)=1} \ket{x}\bra{x}$
denotes the projection onto the subspace spanned by the 
good basis states, 
and similarly
let $\ket{\Psi_0} = {\mathbf P}_{\text{bad}} \ket{\Psi}$
where ${\mathbf P}_{\text{bad}} = \sum_{x:\chi(x)=0} \ket{x}\bra{x}$.
Let $a = \normsqr{\Psi_1}$ denote the probability that a measurement 
of $\ket{\Psi}={\mathcal A}\ket{0}$ yields a good state,
and let $b = \normsqr{\Psi_0}$ denote the probability 
that a measurement of $\ket{\Psi}$ yields a bad state.
We~then have that 
$\ket{\Psi} = \ket{\Psi_1} + \ket{\Psi_0}$ 
and $1 = a+b$.
Finally, let angle~$\theta$ be so that 
$0 \leq \theta \leq \pi/2$ and $a = \sin^2(\theta)$.

As~an example,
for $N =2^n$, we~obtain Grover's searching algorithm~\cite{Grover97}
by setting ${\mathcal A}$ to be the Walsh-Hadamard transform 
on $n$~qubits, letting $\chi(x)$ be~1 
if and only if the database holds a~1 at position~$x$,
and picking phases \mbox{$\phi = \varphi = \pi$}.
If~the database contains a~1 at~$t$ different positions
then $a = t/N$.

The operator ${\mathbf Q}$ implements a unitary operation
on the subspace spanned by $\ket{\Psi_1}$ and~$\ket{\Psi_0}$.
This subspace has dimension~2 if $0<a<1$.
With respect to the ordered orthonormal basis
$\boldsymbol{\big(}\frac{1}{\sqrt{a}}\ket{\Psi_1}, 
  \frac{1}{\sqrt{b}}\ket{\Psi_0}\boldsymbol{\big)}$, 
we can represent
$\mathbf Q = {\mathbf Q}({\mathcal A},\chi,\phi,\varphi)$
by the $2 \times 2$ unitary matrix
\begin{equation}
\label{eq:M}
M = 
\begin{bmatrix}
   -\{(1-e^{\imath \phi})a+e^{\imath \phi}\}
&  (1-e^{\imath \phi})\sqrt{a}\sqrt{1-a} 
e^{\imath \varphi} \\[.41ex]
(1-e^{\imath \phi})\sqrt{a}\sqrt{1-a}  
& \{(1-e^{\imath \phi})a-1\}e^{\imath \varphi}
\end{bmatrix}.
\end{equation}
If $\phi=\varphi=\pi$, then this simplifies to~\cite{BBHT98}
\begin{equation}\label{eq:Msimpel}
M = \begin{bmatrix}
\cos(2 \theta) & -\sin(2 \theta) \\[.11ex]
\makebox[\coslength][r]{$\sin$}(2 \theta) 
& \makebox[\msinlength][r]{$\cos$}(2 \theta) 
\end{bmatrix}.
\end{equation}
That is, if we pick $\phi =\varphi = \pi$, then
each application of~${\mathbf Q}$ implements 
a rotation by angle~$2 \theta$.
A~natural question then is, 
what happens if at least one of the two
angles $\phi$ and~$\varphi$ is not equal to~$\pi$?

\section{Arbitrary rotations}
\label{sec:arbitrary}
Consider the matrix~$M$ defined by Eq.~(\ref{eq:M}).
Our~primary objective is to ensure that the diagonal   
elements of~$M$ are equal.

\vspace{.17 \baselineskip}
{\flushleft {\bf Theorem~1}}\hspace{.4em}
({\bf Optimal angles})\hspace{.4em}{\it
Suppose $\phi \neq \pi$.
Then the two diagonal elements of~$M$ are equal if and only if 
\begin{equation}\label{eq:equal}
\tan(\varphi/2) = \tan(\phi/2) \smallspace(1-2a)
\end{equation}
where matrix~$M$ is defined by Eq.~(\ref{eq:M}).
}
\vspace{0.35 \baselineskip}

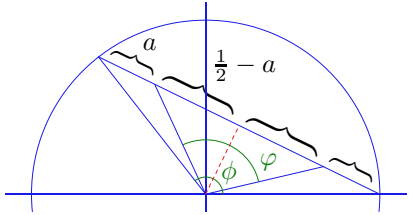
\begin{figure}[htb]
\begin{center}
\setlength{\unitlength}{2160sp}
\begin{picture}(4400,3200)(600,500)
\put(890,2412){\begin{turn}{64}{$\Big\}$}\end{turn}}
\put(1280,2655){$a$}
\put(3395,1191){\begin{turn}{64}{$\Big\}$}\end{turn}}
\put(1478,2040){\begin{turn}{64}{\boldmath${\Bigg\}}$}\end{turn}}
\put(2080,2405){$\frac12-a$}
\put(2440,1560){\begin{turn}{64}{\boldmath${\Bigg\}}$}\end{turn}}
\put(2000,1000){}
\put(2000,1000){\arc{4000}{-3.24159}{0.1}}
\put(-300,1000){\line(1,0){4600}}
\put(2000,800){\line(0,1){2400}}
\path(768.677,2576.02)(4000,1000)
\path(768.677,2576.02)(2000,1000)
\path(1414.94,2260.82)(2000,1000)
\path(3353.74,1315.20)(2000,1000)
\put(2000,1000){}
\put(2000,1000){\arc{400}{-2.2340}{0}}
\put(2000,1000){\arc{1250}{-2.0052}{-.2287}}
\put(2180,1180){$\phi$}
\put(2620,1330){$\varphi$}
\put(2000,1000){}
\path(2000,1000)(2019.22,1039.40)
\path(2038.43,1078.80)(2057.65,1118.20)
\path(2076.87,1157.60)(2096.08,1197.00)
\path(2115.30,1236.40)(2134.52,1275.80)
\path(2153.74,1315.20)(2172.95,1354.60)
\path(2192.17,1394.01)(2211.39,1433.41)
\path(2230.60,1472.81)(2249.82,1512.21)
\path(2269.04,1551.61)(2288.25,1591.01)
\path(2307.47,1630.41)(2326.69,1669.81)
\path(2345.90,1709.21)(2365.12,1748.61)
\put(2000,1000){}
\end{picture}
\caption{Angle $\varphi$ as a function of $\phi$ and~$a$.
Let $\ell$ be the length of the dotted line.
Then $\tan(\varphi/2) = \big(\frac12 -a\big) \frac{1}{\ell}$ 
and $\tan(\phi/2) = \frac12 \frac{1}{\ell}$, 
and hence $\tan(\varphi/2) = \tan(\phi/2) (1-2a)$.}
\label{fig:diagonal}
\end{center}
\end{figure}

Equation~(\ref{eq:equal}) expresses the phase condition
we want to impose on $\phi$ and~$\varphi$.
Theorem~1 can be proven straightforwardly
using standard identities of the trigonometric functions.
Figure~\ref{fig:diagonal} illustrates the proof for the
case that $0 \leq \phi <\pi$ and $0\leq a \leq\frac12$.
The other cases are proven similarly.

We~now show how to implement arbitrary rotations
(up to certain phase factors) by appropriate choices 
of $\phi$ and~$\varphi$.
Let $0 \leq \vartheta < 2 \pi$ be any angle for which
$|\sin(\vartheta)| \leq \sin(2 \theta)$.
First we pick angle $\phi$ so that the absolute value of
the lower left entry of~$M$ equals~$|\sin(\vartheta)|$.
Then we pick angle $\varphi$ so 
that Eq.~(\ref{eq:equal}) holds. 
This ensures that the two diagonal elements of~$M$ are equal.
With these choices of $\phi$ and~$\varphi$,  
matrix~$M$ can thus be written in the form
\begin{equation}
\label{eq:H}
H \,=\,
e^{\imath v}
\begin{bmatrix}
1 & \\[.11ex] & e^{\imath u}
\end{bmatrix}
\begin{bmatrix}
\cos(\vartheta) & -\sin(\vartheta) \\[.11ex]
\makebox[\coslength][r]{$\sin$}(\vartheta) 
& \makebox[\msinlength][r]{$\cos$}(\vartheta) 
\end{bmatrix}
\begin{bmatrix}
1& \\[.11ex] & e^{-\imath u}
\end{bmatrix}
\end{equation}
for some angles $0 \leq u,v < 2 \pi$.
We~denote this matrix by~$H$.
Here and elsewhere,
missing matrix entries are assumed equal to~0.

{To}~summarize, 
we~have just shown that for all $a$ ($0<a<1$)
and for all angles $\vartheta$ ($0 \leq \vartheta <2\pi$)
so that $|\sin(\vartheta)| \leq \sin(2\theta)$
there exist angles $0\leq \phi,\varphi < 2\pi$ 
so that $M=H$ for some  angles $0 \leq u,v<2\pi$, 
where $M$ is given by Eq.~(\ref{eq:M}) and
$H$ by Eq.~(\ref{eq:H}).
Matrix~$H$ represents a rotation 
by angle~$\vartheta$ which is conjugated by a 
conditional phase change by angle~$u$, up to 
a global phase factor of~$e^{\imath v}$.

In~many applications, applying~$H$ will be equally good
to applying a rotation by angle~$\vartheta$.
For instance, we can use~$H$ to implement arbitrary rotations
as follows.
Suppose $a$ is known. 
Let $x$ be any angle ($0 \leq x < 2\pi$).
We~implement a rotation by angle~$x$ as follows.
First, we check if~$x$ is a multiple of~$2 \theta$.
If~so, we simply just apply
${\mathbf Q}({\mathcal A},\chi,\pi,\pi)$ 
a total number of $x/(2\theta)$ times and stop~\cite{BBHT98}.
Otherwise, we compute the smallest integer~$m$ larger than
$x/(2\theta)$ and we set $\vartheta = x/m$. 
Then we find angles $\phi$ and $\varphi$ so that $M=H$,
and we compute the angles $u$ and~$v$.
With these choices of angles, we can factorize
the rotation by angle~$x$ as 
\begin{equation}
\begin{bmatrix}
\cos(x) & -\sin(x) \\[.11ex]
\makebox[\coslength][r]{$\sin$}(x) 
& \makebox[\msinlength][r]{$\cos$}(x) 
\end{bmatrix}
\,=\, e^{-\imath mv}
\begin{bmatrix}
1 & \\[.11ex] & e^{-\imath u}
\end{bmatrix}
M^m
\begin{bmatrix}
1& \\[.11ex] & e^{\imath u}
\end{bmatrix}.
\end{equation}
Thus, to implement a rotation by angle~$x$, we
first apply a conditional phase-change of the bad
states by angle~$u$.  Then we apply
${\mathbf Q}({\mathcal A},\chi,\phi,\varphi)$ 
a total number of $m$ times, apply a conditional 
phase-change of the bad states by angle~$-u$, 
and finally apply a global phase-change by angle~$-mv$.

\section{Obtaining success probability~1}
\label{sec:certainty}
We~now show that if we pick nontrivial angles 
$\phi$ and~$\varphi$ so that Eq.~(\ref{eq:equal}) holds,
then we can find a good solution with certainty.
That is, we can use any set of angles $(\phi,\varphi) \neq (0,0)$ 
for which Eq.~(\ref{eq:equal}) holds.

Let angle $0 \leq \vartheta\leq \pi/2$ be defined 
so that $\sin(\vartheta) = |\sin(\phi/2) \sin(2\theta)|$.
Since $\phi$ and~$\varphi$ satisfy Eq.~(\ref{eq:equal}), 
we~can write ${\mathbf Q}({\mathcal A},\chi,\phi,\varphi)$ 
in the form
\begin{equation}
e^{\imath v}
\begin{bmatrix}
1 & \\[.11ex] & e^{\imath u}
\end{bmatrix}
\begin{bmatrix}
\cos(\vartheta) & -\sin(\vartheta) \\[.11ex]
\makebox[\coslength][r]{$\sin$}(\vartheta) 
& \makebox[\msinlength][r]{$\cos$}(\vartheta) 
\end{bmatrix}
\begin{bmatrix}
1& \\[.11ex] & e^{-\imath u}
\end{bmatrix}
\end{equation}
for some angles $0 \leq u,v < 2 \pi$.
That is, operator~${\mathbf Q}({\mathcal A},\chi,\phi,\varphi)$ 
implements a rotation by angle~$\vartheta$, up to phase factors.

Our idea for finding a good solution with certainty
is as follows: 
Let $m = \big\lceil \frac{\pi/2-\theta}{\vartheta}\big\rceil$
and $\theta_{\text{init}} = \frac{\pi}{2} - m \vartheta$.
Then $-\theta < \theta_{\text{init}} \leq \theta$.
We~first set up a superposition representing the initial
angle~$\theta_{\text{init}}$.  This is possible since 
$|\theta_{\text{init}}| \leq \theta$.
Then we apply operator~${\mathbf Q}({\mathcal A},\chi,\phi,\varphi)$
a total number of~$m$ times and finally we measure.
This produces a good solution with certainty.

A~realization of this idea is as follows:
Let $\ket{\Psi_{\text{init}}}$ denote the state
$\sin(\theta_{\text{init}}) \frac{1}{\sqrt{a}} \ket{\Psi_1}
+\cos(\theta_{\text{init}}) \frac{1}{\sqrt{1-a}} 
\ket{\Psi_0}$.
First we apply $\mathcal A$ on the first register
of the state~$\ket{0}\ket{0}\ket{0}$, producing
the state~$\ket{\Psi}\ket{0}\ket{0}$.
Then we unitarily map $\ket{\Psi}\ket{0}\ket{0}$ to the state
$\alpha \ket{\Psi_{\text{init}}}\ket{0}\ket{0}
+ \beta \ket{0}\ket{\Psi_1}\ket{1}$ for 
some complex numbers $\alpha, \beta \in \Co$.
Then we apply operator~${\mathbf Q}({\mathcal A},\chi,\phi,\varphi)$
a total number of~$m$ times on the first register,
producing the state
$\alpha' \ket{\Psi_1}\ket{0}\ket{0}
+ \beta' \ket{E}\ket{\Psi_1}\ket{1}$ for 
some complex numbers $\alpha', \beta' \in \Co$
and some state~$\ket{E}$.
Finally, we swap the contents of the first two registers
conditionally to that the third register contains a~1, producing the
final tensor product state $\ket{\Psi_1}\ket{E'}_{2,3}$,
for some state~$\ket{E'}_{2,3}$ that represents the joint
state of registers~2 and~3.

We~may summarize this section by saying that for fixed
known rotational angle~$\vartheta$, 
we can modify the angle of the initial state 
so that we succeed with certainty.

\section{The phase condition}
\label{sec:phasecondition}
Our condition that $\tan(\varphi/2) = \tan(\phi/2) (1-2a)$
implies unfortunately that $\phi$ and~$\varphi$ depend 
nontrivially on~$a$.
Put formally, for all angles $0<\phi<2\pi$ so 
that $\phi\neq\pi$, the following holds:
For all angles $0\leq\varphi<2\pi$ with $\varphi\neq\pi$,
there exists a unique
$a \in \Re$ so that Eq.~(\ref{eq:equal}) holds,
and for all $0\leq a\leq1$, there exists a unique angle
$0\leq\varphi<2\pi$ so that Eq.~(\ref{eq:equal}) holds.
If~the success probability~$a$ is known in advance, then for 
any angle $\phi$ we may want to pick, we can easily compute
the angle $\varphi$ to use.  
However, if $a$ is not known, then this is not possible
unless $\phi \in \{0,\pi\}$.  Thus for 
angles $\phi \not\in\{0,\pi\}$,
we need to know~$a$ to compute~$\varphi$ so that
Eq.~(\ref{eq:equal}) holds.

If~we do not know $a$ in advance, then a subsidiary strategy 
could be to utilize a set of angles $(\phi,\varphi)$ so that 
Eq.~(\ref{eq:equal}) almost holds.
If~$a$ is small then we could for example 
approximate $\varphi$ by~$\phi$.  
In~the papers~\cite{LZLN99,LLZN99,LTLZY99,LLZT00}, 
Long~\emph{et~al.} consider the question of when 
arbitrary phases can be utilized successfully in quantum searching. 
Their conclusion is that the angles have to 
equal ($\phi = \varphi$) ``to construct 
an efficient quantum search algorithm''~\cite{equal}.
Our condition that $\tan(\varphi/2) = \tan(\phi/2) (1-2a)$
is obviously different from their condition 
(that $\phi = \varphi$)
whenever $0 < a< 1$ and $\phi \not\in \{0,\pi\}$.
The explanation for these different results is that 
Long~\emph{et~al.} consider when the 
quantum search algorithm succeeds with high probability,
whereas we, in the previous section, consider when the 
quantum search algorithm succeeds with certainty.
In~particular, all our calculations are exact.
A~main proof technical idea used by 
Long \emph{et~al.} for example in~\cite{LLZN99}
is approximations of the type $K_1 \approx e^{K_2}$
for $2 \times 2$ matrices $K_1$ and~$K_2$.

Long \emph{et~al.}~\cite{LLZN99,LTLZY99} prove their result 
via an ${\mathbf SO}(3)$ rotational interpretation of
operator~${\mathbf Q}({\mathcal A},\chi,\phi,\varphi)$.
We~now reprove the theorem of Long~\emph{et~al.} that 
if we use phases $\phi=\varphi$, then we can find a good
solution with high probability.
The main idea in our alternative proof is to consider the case
$\phi=\varphi$ an approximation to the perfect case in 
which Eq.~(\ref{eq:equal}) holds, and then lower bound how
well this approximation works.
We~do that by upper bounding the norm of the difference 
of two operators.
This idea provides a short and straightforward proof.

\vspace{.17 \baselineskip}
{\flushleft {\bf Lemma~1}}\hspace{.4em}
({\bf Equal angles})\hspace{.4em}{\it
Let $0 < \phi <\pi$.
Let angle $0 < \vartheta\leq \pi/2$ be so that 
$\sin(\vartheta) = \sin(\phi/2)\sin(2\theta)$.  Let 
$m = \big\lceil \frac{\pi/2}{\vartheta} -\frac12\big\rceil$
and let ${\mathbf Q}' = {\mathbf Q}({\mathcal A},\chi,\phi,\phi)$.
Then
\begin{equation}
| \langle \Psi'_1 |  {{\mathbf Q}'}^{m} {\mathcal A} |0\rangle| 
\;\geq\; 1 - a (2+4\pi^2m)
\end{equation}
where $\ket{\Psi'_1} = \frac{1}{\sqrt{a}}\ket{\Psi_1}$.}
\vspace{0.35 \baselineskip}

Let angle $-\pi < \varphi<\pi$ be so that
$\tan(\varphi/2) = \tan(\phi/2) (1-2a)$.
Then $|\phi-\varphi| \leq 2 \pi a$.
Let ${\mathbf Q} = {\mathbf Q}({\mathcal A},\chi,\phi,\varphi)$
and let $\mathcal H$ denote the Hilbert space that ${\mathbf Q}$ 
and ${\mathbf Q}'$ act upon.
Let $\|\cdot\|$ denote the operator norm on~$\mathcal H$ defined 
by $\|O\| = \sup\{ |O\ket{\Gamma}| \;:\; 
|\smallspace\ket{\Gamma}\smallspace|=1\}$
for any operator~$O$ on~$\mathcal H$.
Then 
\begin{equation}\label{eq:norm}
\big\| {\mathbf Q}' - {\mathbf Q}\big\| =
\bigg\| \begin{bmatrix}1&\\&e^{\imath \phi}\end{bmatrix}
    - \begin{bmatrix}1&\\&e^{\imath \varphi}\end{bmatrix} \bigg\|
= \big|1 - e^{\imath (\phi-\varphi)}\big|
\leq 4 \pi^2 a,
\end{equation}
and thus
$\big\| {{\mathbf Q}'}^m - {\mathbf Q}^m\big\| 
\leq 4 \pi^2 a m$.

By~definition of~$m$ we have that 
$\big|\frac{\pi}{2}  -m \vartheta\big| 
\leq \frac{\vartheta}{2} \leq \theta$,
so $\sin(m\vartheta) \geq \sqrt{1-a}$
and hence
$\big| \langle \Psi'_1 |  {\mathbf Q}^{m} 
|\Psi\rangle \big| \geq 1-2a$.
Thus,
\begin{equation}
\big| \langle \Psi'_1 |  {{\mathbf Q}'}^{m} 
|\Psi\rangle \big| 
\;\geq\; \big| \langle \Psi'_1 |  {\mathbf Q}^{m} 
  |\Psi\rangle \big| -4\pi^2 a m
\;\geq\; 1 - a (2+4\pi^2m).
\end{equation}
Lemma~1 follows.

If~we measure the state
${{\mathbf Q}'}^{m} {\mathcal A} |0\rangle$,
then the outcome is good with probability 
at least $1-4a(1+2\pi^2 m)$, which is at least
$1-4a(\frac{\pi^3}{\vartheta}+11)$.
The probability of measuring a bad state is thus upper bounded 
by $4\pi^3\frac{a}{\vartheta}+44a$,
which is $O\big(\frac{\sqrt{a}}{\phi}\big)$
for $0<a\leq 1$ and $0 < \phi<\pi$.
Theorem~2 follows.

\vspace{.17 \baselineskip}
{\flushleft {\bf Theorem~2}}\hspace{.4em}
({\bf Equal angles})\hspace{.4em}{\it
Let $\mathcal A$ be any quantum algorithm that uses no measurements.
Let $a$ denote the success probability of~$\mathcal A$ 
and let angle $0 < \theta \leq \pi/2$ 
be so that $\sin^2(\theta)=a$.
Let $\phi$ be any angle so that  $\theta \leq \phi <\pi$.
Let $m = \big\lceil \frac{\pi/2}{\vartheta} -\frac12\big\rceil$
where angle $0 < \vartheta\leq \pi/2$ is so that 
$sin(\vartheta) = \sin(\phi/2) \sin(2\theta)$.
Let ${\mathbf Q}' = {\mathbf Q}({\mathcal A},\chi,\phi,\phi)$.
Then a measurement of 
${{\mathbf Q}'}^{m} {\mathcal A} |0\rangle$
will provide a good solution with 
probability~$1-O\big(\frac{1}{\phi}\sqrt{a}\big)$.}
\vspace{0.35 \baselineskip}

Theorem~2 relies on two properties:
Firstly, that operator 
${\mathbf Q}({\mathcal A},\chi,\phi,\phi)$
approximates operator
${\mathbf Q} = {\mathbf Q}({\mathcal A},\chi,\phi,\varphi)$
sufficiently well.
Secondly, that each application of~${\mathbf Q}$
implements a rotation by a sufficiently large angle~$\vartheta$.
Any set of angles $(\phi,\varphi)$ 
so that these two properties are fulfilled
will provide a quantum amplitude amplification scheme.
In~general, the closer we pick the rotational angle~$\vartheta$ 
to the maximal angle~$2\theta$,
the worse approxi\-ma\-tion for~${\mathbf Q}$ can we allow 
ourselves to use, and vice~versa.
In~the next theorem, which is proven almost identically
to Theorem~2, 
we express one way of capturing this duality.

\vspace{.17 \baselineskip}
{\flushleft {\bf Theorem~3}}\hspace{.4em}
({\bf Any angles})\hspace{.4em}{\it
Let $\mathcal A$ be any quantum algorithm that uses no measurements.
Let $a$ denote the success probability of~$\mathcal A$ 
and let angle $0 < \theta \leq \pi/2$ be so that $\sin^2(\theta)=a$.
Let $0< \phi<\pi$ and
$-\pi< \varphi'<\pi$ be given angles.

Let $m = \big\lceil \frac{\pi/2}{\vartheta} -\frac12\big\rceil$
where angle $0 < \vartheta\leq \pi/2$ is so that 
$sin(\vartheta) = \sin(\phi/2) \sin(2\theta)$.
Let $-\pi<\varphi<\pi$ be defined so that 
$\tan(\varphi/2) = \tan(\phi/2)(1-2a)$ and 
let $\delta = |\varphi'-\varphi|$.
Let ${\mathbf Q}' = {\mathbf Q}({\mathcal A},\chi,\phi,\varphi')$
denote our approximation to 
${\mathbf Q}({\mathcal A},\chi,\phi,\varphi)$.
Then a measurement of ${{\mathbf Q}'}^{m} {\mathcal A} |0\rangle$
will provide a good solution with 
error probability at most $4a+\epsilon$,
provided 
$\delta \leq \epsilon\smallspace 
\frac{\sqrt{3}}{2\pi^2(\sqrt{3}+\pi)} \smallspace \phi\sqrt{a}$.}
\vspace{0.35 \baselineskip}

The above theorem put bounds on 
the error $\delta=|\varphi'-\varphi|$ that
we can tolerate to still obtaining a quantum algorithm 
that succeeds with high probability.
Suppose $\phi$ is a constant, say~$\phi=\frac{\pi}{10}$,
then whenever $\delta$ is at most $c \sqrt{a}$, for some 
appropriate constant~$c$, the above algorithm finds a 
good solution with bounded error probability.
Furthermore,
if~$\delta$ is at most~$O(a)$, as it is if we pick 
$\varphi'=\phi$, 
then the error probability drops to being in~$O(\sqrt{a})$.
Thus, the smaller error in the choice of angles from
the perfect case as expressed by Eq.~(\ref{eq:equal}),
the smaller is the error probability of the algorithm.

Theorem~3 is similar 
to the main result in~\cite{LLZT00} where it is proven that for quantum
searching, the distance $|\phi-\varphi|$
must be at most of order~$\frac{1}{\sqrt{N}}$ 
for finding the marked element with constant success probability.
Our result generalizes their result as it measures the
error probability of the overall algorithm in terms of the 
distance from the perfect case in which Eq.~(\ref{eq:equal})
holds.  The case $\phi=\varphi$ is already an approximation,
which by itself introduces an error in~$\delta$ 
in the order of~$\Theta(a)$.
In~addition, Theorem~3 includes the cases that the 
angles~$\phi$ and $\varphi$ are not constants but depend on~$a$.
However, the main benefit of Theorem~3
is that it is so easy to prove when one is given Theorem~1.
Essentially, the proof of Theorem~3 
reduces to bounding the distance between the two 
operators~${\mathbf Q}({\mathcal A},\chi,\phi,\varphi)$ 
and~${\mathbf Q}({\mathcal A},\chi,\phi,\varphi')$,
which is easy.

\section{Conclusion}
\label{sec:conclusion}
Amplitude amplification is besides quantum Fourier 
transforms the most successfully used tool in quantum algorithms. 
It~is a generalization of Grover's quantum 
searching algorithm 
and it allows a quadratic speed up of many algorithms.

We~prove that the phase condition in amplitude amplification 
can be expressed by the equation
$\tan(\varphi/2) = \tan(\phi/2) (1-2a)$,
where $a$ denotes the success probability of the original algorithm.
We~show how to implement arbitrary rotations 
and how to boost quantum algorithms to succeed with certainty
by utilizing angles $(\phi,\varphi)$ that satisfy 
the above equation.
In~both cases, the number of iterations required 
increases linearly in the inverse of angle~$\phi$.
For instance, if we choose $\phi = \frac{\pi}{c}$ 
for some constant~$c>1$,
then we require $\Theta\big(c \frac{1}{\sqrt{a}}\big)$
iterations to find a solution with certainty.

Whenever the success probability~$a$ is not known a~priori, we 
can approximate the above trigonometric equation by the 
linear equation $\phi=\varphi$.  
This case has been studied by 
Long \emph{et~al.}~\cite{LZLN99,LLZN99,LTLZY99,LLZT00}
and in particular, they have shown that equal angles can
be utilized in quantum searching.
By~considering the case $\phi=\varphi$ an approximation
to the perfect case $\tan(\varphi/2) = \tan(\phi/2) (1-2a)$,
we can reprove the main results by Long \emph{et~al.}
We~believe that this approach makes the analysis and proofs 
straightforward and easy to understand.

\section*{Acknowledgements}
I~am grateful to Gilles Brassard and Richard Cleve for comments.



\begin{thebibliography}{8}

\let\bib\bibitem
        \bib{Grover97}
L.\,K.{} Grover, 
``Quantum mechanics helps in searching for a needle in a haystack,''
{Phys.{} Rev.{} Lett.}
{\textbf 79}, 325 (1997).

\let\bib\bibitem
        \bib{LZLN99}
G.\,L.{} Long, W.\,L.{} Zhang, Y.\,S.{} Li, and L.~Niu,
``Arbitrary phase rotation of the marked state
 cannot be used for {Grover's} quantum search algorithm''
(Rapid Communications),
{Commun.{} Theor.{} Phys.}
{\textbf 32}(3), 335 (1999).

\let\bib\bibitem
        \bib{LLZN99}
G.\,L.{} Long, Y.\,S.{} Li, W.\,L.{} Zhang, and L.~Niu,
``Phase matching in quantum searching,''
{Phys.{} Lett.{} A} {\textbf 262}(1), 27
(1999).

\let\bib\bibitem
        \bib{LTLZY99}
G.\,L.{} Long, C.\,C.{} Tu, Y.\,S.{} Li, W.\,L.{} Zhang, 
and H.\,Y.{} Yan,
``A~novel \mbox{SO(3)} picture for quantum searching,''
Available at Los Alamos e-Print archive 
as \myurl{arXiv.org/abs/quant-ph/9911004}.

\let\bib\bibitem
        \bib{LLZT00}
G.\,L.{} Long, Y.\,S.{} Li, W.\,L.{} Zhang, and C.\,C.{} Tu,
``Dominant gate imperfection in {Grover's} 
quantum search algorithm,''
{Phys.{} Rev.{} A} {\textbf 61}(4), 042305 (2000).

\let\bib\bibitem
        \bib{BHMT00}
G.~Brassard, P.~H{\o}yer, M.~Mosca, and A.~Tapp,
``Quantum amplitude amplification and estimation.''
Available at Los Alamos e-Print archive 
as \myurl{arXiv.org/abs/quant-ph/0005055}.

\let\bib\bibitem
        \bib{BBHT98}
M.~Boyer, G.~Brassard, P.~H{\o}yer, and A.~Tapp,
``Tight bounds on quantum searching,''
{Fortschr.{} Phys.}
{\textbf 46}, 493 (1998).

\let\bib\bibitem
	\bib{equal}
See for example the first paragraph on page 042305-2 
in~\cite{LLZT00}.
Please note that the notation used 
in~\cite{LZLN99,LLZN99,LTLZY99,LLZT00}
is slightly different from ours: 
their~$\theta$ denotes our~$\varphi$,
and their~$\varphi$ denotes our~$\phi$.

\end{thebibliography}
\end{document}